\documentclass[12pt,preprint]{aastex}

\begin{document}
\title{Characteristics of EGRET Blazars in the VLBA Imaging and Polarimetry Survey (VIPS)}
\author{G. B. Taylor\altaffilmark{1}, S. E. Healey\altaffilmark{2}, J. F. Helmboldt\altaffilmark{3},  S. Tremblay\altaffilmark{1}, C. D. Fassnacht\altaffilmark{4}, R. C. Walker\altaffilmark{5}, L. O. Sjouwerman\altaffilmark{5}, T. J. Pearson\altaffilmark{6}, A. C. S. Readhead\altaffilmark{6}, L. Weintraub\altaffilmark{6}, N. Gehrels\altaffilmark{7}, R. W. Romani\altaffilmark{2},  P. F. Michelson\altaffilmark{2}, R. D. Blandford\altaffilmark{8}, and G. Cotter\altaffilmark{9}}

\altaffiltext{1}{Department of Physics and Astronomy, University of New Mexico, 800 Yale Blvd NE, Albuquerque, NM 87131, USA}
\altaffiltext{2}{Department of Physics, Stanford University, Stanford, CA 94305}
\altaffiltext{3}{Naval Research Laboratory, Washington, DC 20375}
\altaffiltext{4}{Department of Physics, University of California at Davis, 1 Shields Avenue, Davis, CA 95616}
\altaffiltext{5}{National Radio Astronomy Observatory, P.O. Box O, Socorro, NM 87801, U.S.A.}
\altaffiltext{6}{Astronomy Department, California Institute of Technology, Mail Code 105-24, 1200 East California Boulevard, Pasadena, CA 91125}
\altaffiltext{7}{NASA Goddard Space Flight Center, Greenbelt, MD 20771}
\altaffiltext{8}{KIPAC, Stanford University, PO Box 20450, MS 29, Stanford, CA 94309, USA}
\altaffiltext{9}{University of Oxford, Department of Astrophysics, Denys Wilkinson Building, Keble Road, Oxford OX1 3RH}
\received{?}
\accepted{?}

\begin{abstract}

We examine the radio properties of EGRET-detected blazars observed as
part of the VLBA Imaging and Polarimetry Survey (VIPS).  VIPS has a
flux limit roughly an order of magnitude below the MOJAVE survey and
most other samples that have been used to study the properties of
EGRET blazars.  At lower flux levels, radio flux density does not
directly correlate with gamma-ray flux density.  We do find that the
EGRET-detected blazars tend to have higher brightness temperatures,
greater core fractions, and possibly larger than average jet opening
angles.  A weak correlation is also found with jet length and with
polarization.  All of the well-established trends can be explained by
systematically larger Doppler factors in the $\gamma$-ray loud 
blazars, consistent with the
measurements of higher apparent velocities found in monitoring
programs carried out at radio frequencies above 10 GHz.

\end{abstract}


\section{Introduction}

At high galactic latitudes at least $\sim$70\% of the sources seen by the
Compton Gamma-Ray Observatory (CGRO; Hartman et al. 1999) EGRET
instrument are identified with blazars (Sowards-Emmerd et al. 2003,
2004, 2005).  These blazars exhibit flat radio spectra, rapid variability,
compact cores with one-sided parsec-scale jets, and superluminal
motion in the jets (Marscher 2006).  At optical wavelengths blazars are
characterized by broad or no emission lines, optically violently
variable behavior, and relatively high linear polarization (Urry \&
Padovani 1995). 

The identified EGRET blazars display a number of interesting radio
properties.  Based on 22 and 37 GHz monitoring of 43 sources,
L\"ahteenm\"aki \& Valtaoja (2003) suggest that the EGRET flares
lagged behind mm flares.  This is in direct contrast to other studies
(e.g., Reich et al. 1993) that claimed enhanced radio emission
following a high level of gamma-ray activity.  Lister \& Homan (2005)
find that EGRET blazars have more strongly polarized jets than 
found in compact objects on average based on
analysis of 26 EGRET sources overlapping with the Monitoring of Jets
in Active galactic nuclei with VLBA Experiments (MOJAVE) VLBI survey.
Kovalev et al. (2005) reported that EGRET blazars in the 15 GHz MOJAVE
survey were more compact than non-EGRET blazars.  EGRET blazars have
also been claimed to have much higher superluminal speeds than other
blazars, and to have gamma-ray flares associated with the ejection of
superluminal radio knots (Jorstad et al. 2001a,b; Kellermann et
al. 2004).  However, in consideration of a larger sample (taken at 2
and 8 GHz) Piner et al. (2007) find no significant difference between
the jet velocities of EGRET and non-EGRET blazars.  In either case it
still is unclear whether the radio flaring, or the ejection of new
components, precedes or follows the gamma-ray flare, or whether there
are in fact two classes of gamma-ray blazars (e.g., Sikora et
al. 2001, 2002), one class with steep gamma-ray spectra and another
class with flat gamma-ray spectra.

We have recently completed the 5 GHz VLBA Imaging and Polarimetry Survey
(VIPS) of 1127 flat-spectrum sources stronger than 85 mJy in the
Northern Cap region of the Sloan Digital Sky Survey (SDSS).  VIPS
(Taylor et al. 2005, Helmboldt et al. 2007a) was carried out between
January and August of 2006 and includes 5 GHz, full polarization
images of 958 sources as well as 147 total intensity images for
sources previously observed as part of the Caltech Jodrell Bank Flat
spectrum survey (CJF).  The small remainder (22) were previously
observed in the VIPS pilot survey and/or in the MOJAVE survey.  Over
99\% of the VIPS sources have been successfully imaged, with just under
1\% being too weak to reliably self-calibrate and image.

The large, uniform sample provided by VIPS will have many uses, including
enabling the intelligent follow-up of gamma-ray blazars as they flare
and are detected by the Gamma-ray Large Area Space Telescope (GLAST; Gehrels 
et al.\ 2001).
GLAST, to be launched in late 2007, will provide a dramatic increase
in the resolution and energy range and over two orders of magnitude
increase in sensitivity over CGRO-EGRET, thereby affording an
unprecedented opportunity for the study of the centers of activity and
jets in blazars in the gamma-ray energy range.

Here we make use of the VIPS survey to examine correlations in the 
parsec-scale jet properties of EGRET blazars.  

\section{Sample Definition}

\begin{figure}
\plottwo{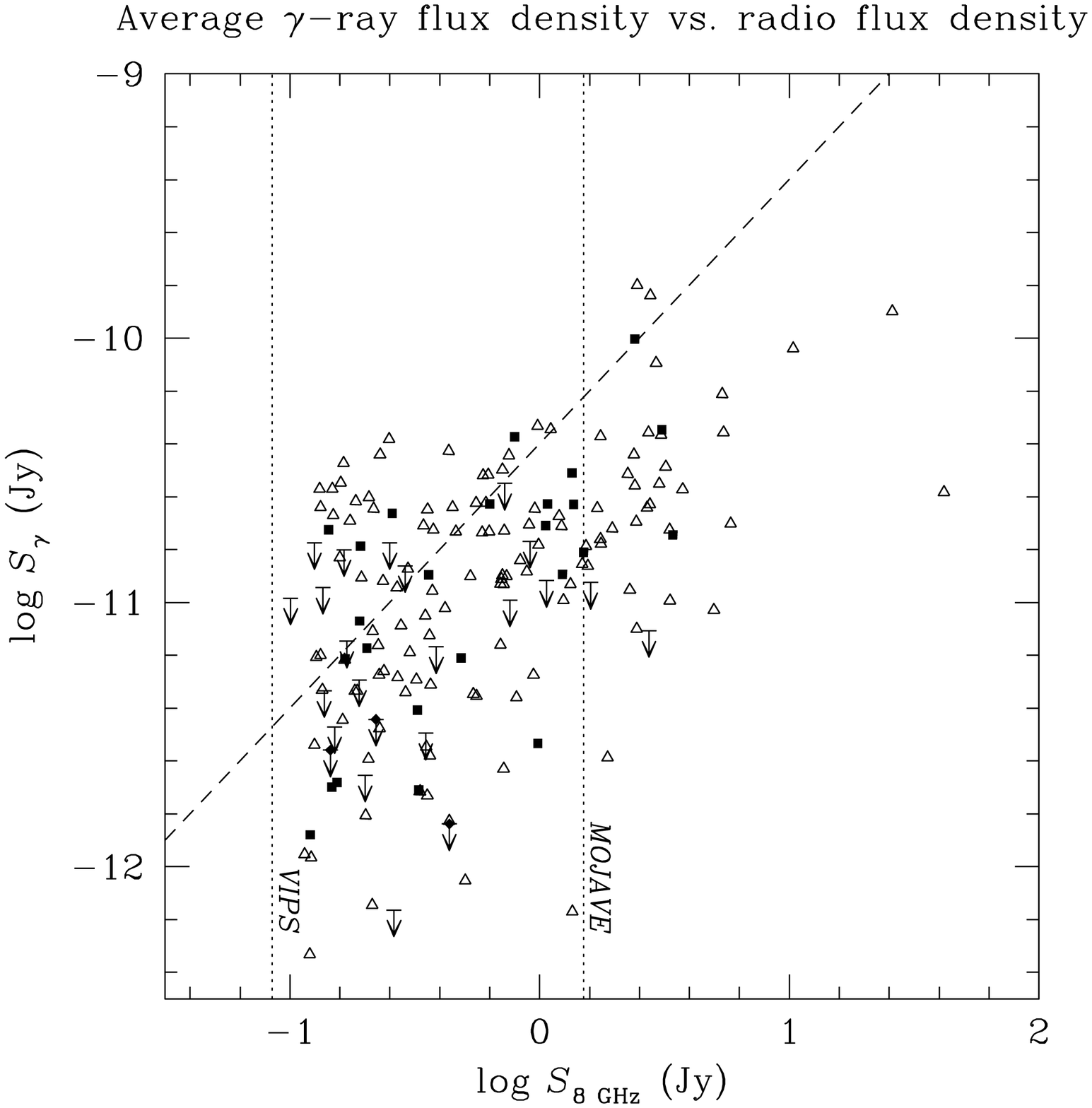}{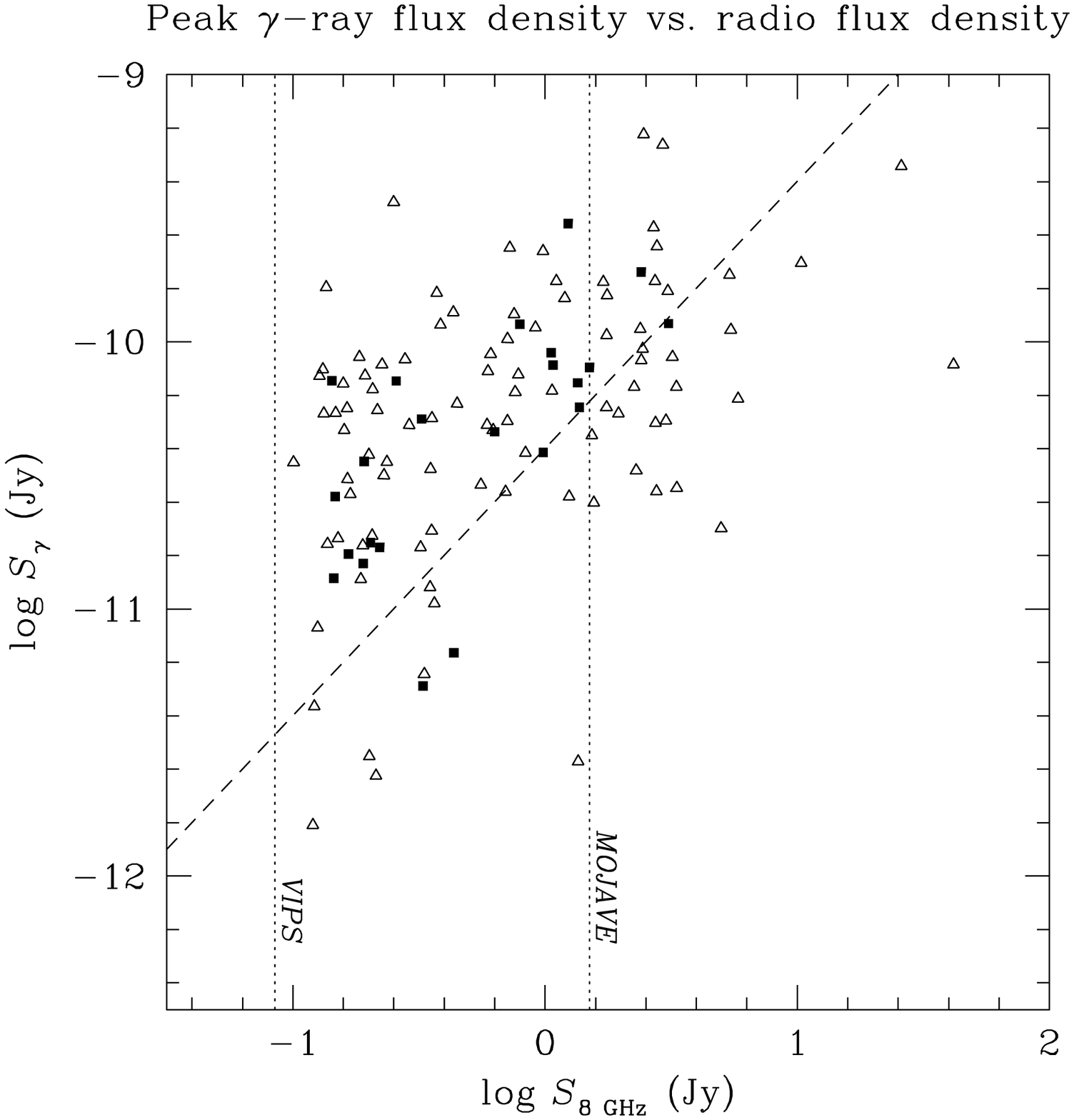}
\caption{{\bf Left:} mean EGRET fluxes  vs.
the 8.4 GHz flux density. The mean fluxes are mission-averaged values 
from all four EGRET observing cycles, listed as ``P1234'' in the 3EG catalog. {\bf Right:}
Peak gamma-ray flux vs. radio flux density, where peak designates the largest
flux from a single (typically two week long) EGRET viewing period, provided the
detection significance exceeded $3\sigma$. The dashed line is not a fit to the
data but is only present to show that assuming $\gamma$-ray
flux proportional to the radio flux does not fit well at low radio flux
densities; it appears that a statistically significant fraction of sub-Jy
blazars may be bright in the $\gamma$-ray. Filled squares indicate VIPS sources
discussed in this paper.  The vertical dashed lines indicate the 
flux density limits of the VIPS and MOJAVE surveys assuming a flat
spectrum (i.e., no correction has been made for spectral index effects).
}
\label{gamma}
\end{figure}

We have compared the latest EGRET catalog (Hartman et al. 1999) to the
recently compiled Combined Radio All-sky Targeted Eight GHz Survey
(CRATES; Healey et al. 2007a), and from the intersection compiled a
sample of 173 candidate EGRET sources at $|b| > 10^\circ$. 
In Fig.~\ref{gamma} we plot the Gamma-ray mean and peak fluxes against
the 8.4 GHz radio flux density from CRATES (Healey et al. 2007a) for 173
and 117 sources respectively.  The VIPS sample
(which covers only $\sim$14\% of the sky), is derived from 
CLASS  (Myers et al. 2003),  but otherwise has similar
radio flux limits and spectral selection as CRATES.  Both CRATES
and CLASS include A configuration VLA observations at 8.4 GHz.
The full
radio/$\gamma$-ray properties of the sample will be discussed in
Healey et al. 2007b.  Here we focus on the candidate EGRET sources
within the $\sim$5700 square degree region of the sky covered by VIPS,
which provides uniform quality VLBI imaging.  The results are summarized in
Table 1.  This list contains 12 VIPS sources deemed likely candidates,
and 19 VIPS sources deemed plausible candidates. Here ``likely'' and
``plausible'' refer to ranges in the Figure of Merit (FoM).  The FoM
is computed based on the 8.4 GHz flux density, the radio spectral
index, the X-ray flux density, and the positional coincidence
(Sowards-Emmerd et al. 2003).  A source is considered ``likely'' if it
has FoM $>$ 1, and ``plausible'' if it has 1.0 $>$ FoM $>$ 0.25.

\begin{figure}
\plotone{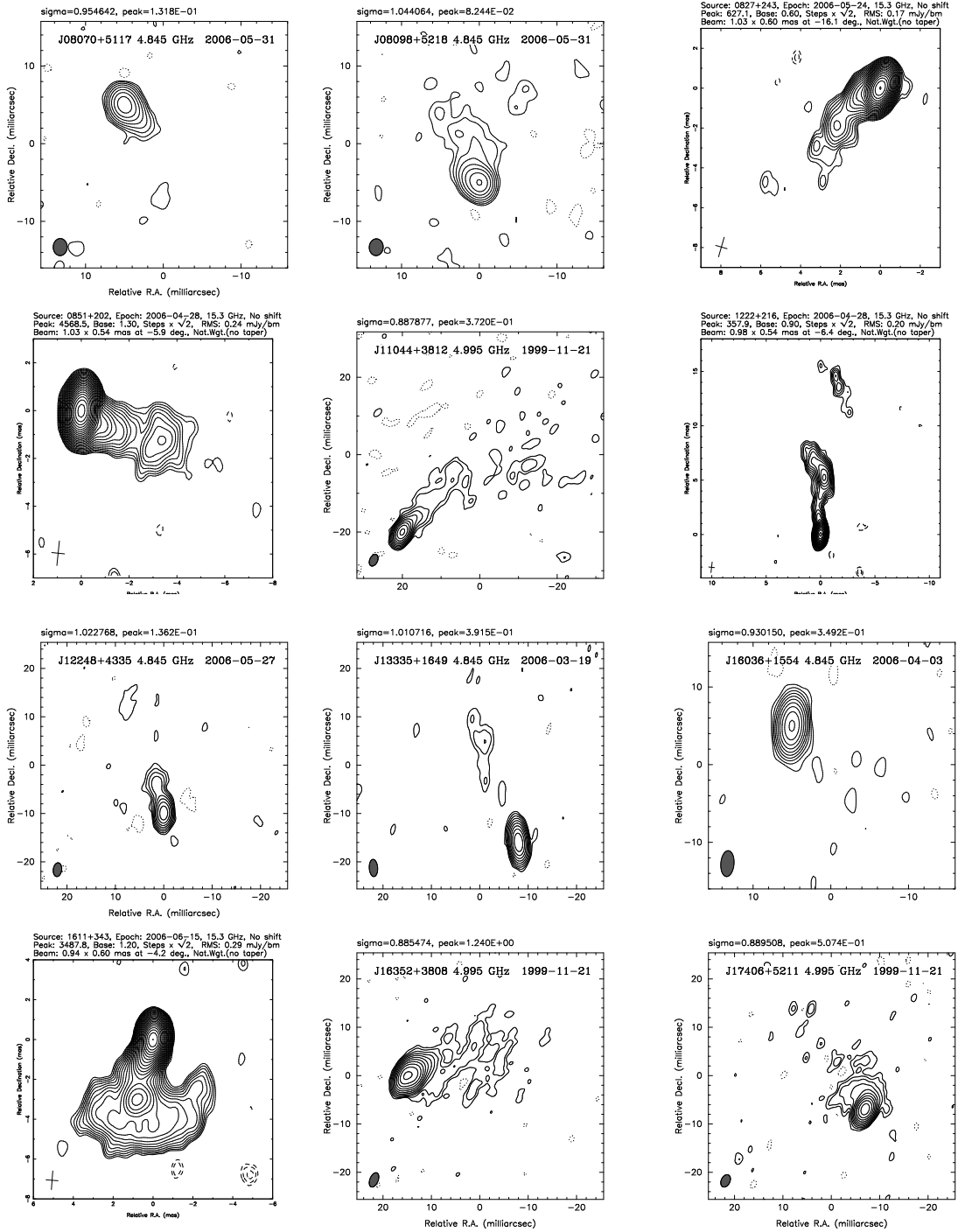}
\caption{Images from VIPS and MOJAVE for 12 sources identified as
likely EGRET candidates.}
\label{likely1}
\end{figure}

\begin{figure}
\plotone{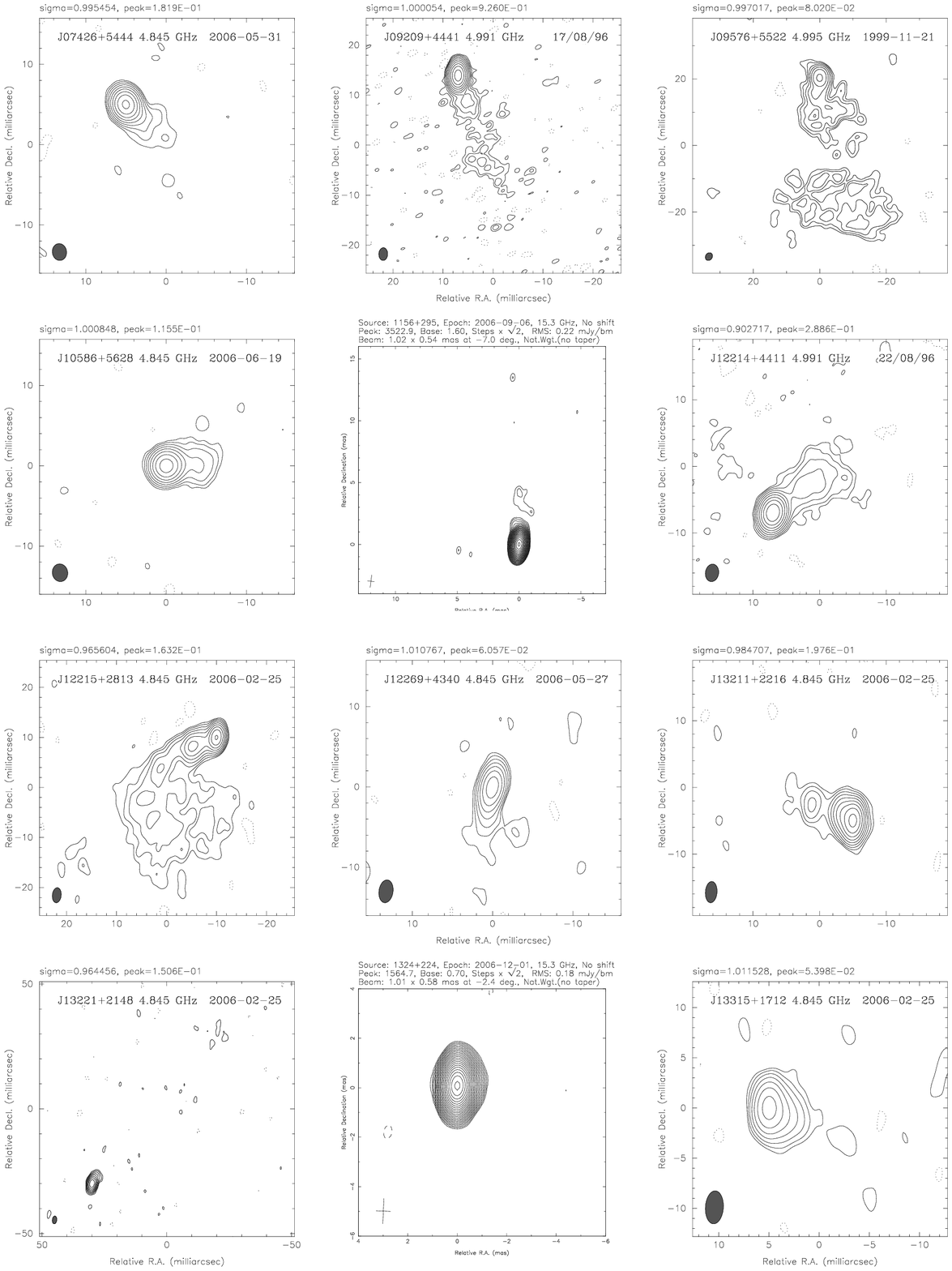}
\caption{Images from VIPS and MOJAVE for 12 sources identified as
plausible EGRET candidates.}
\label{maybe1}
\end{figure}

\begin{figure}
\plotone{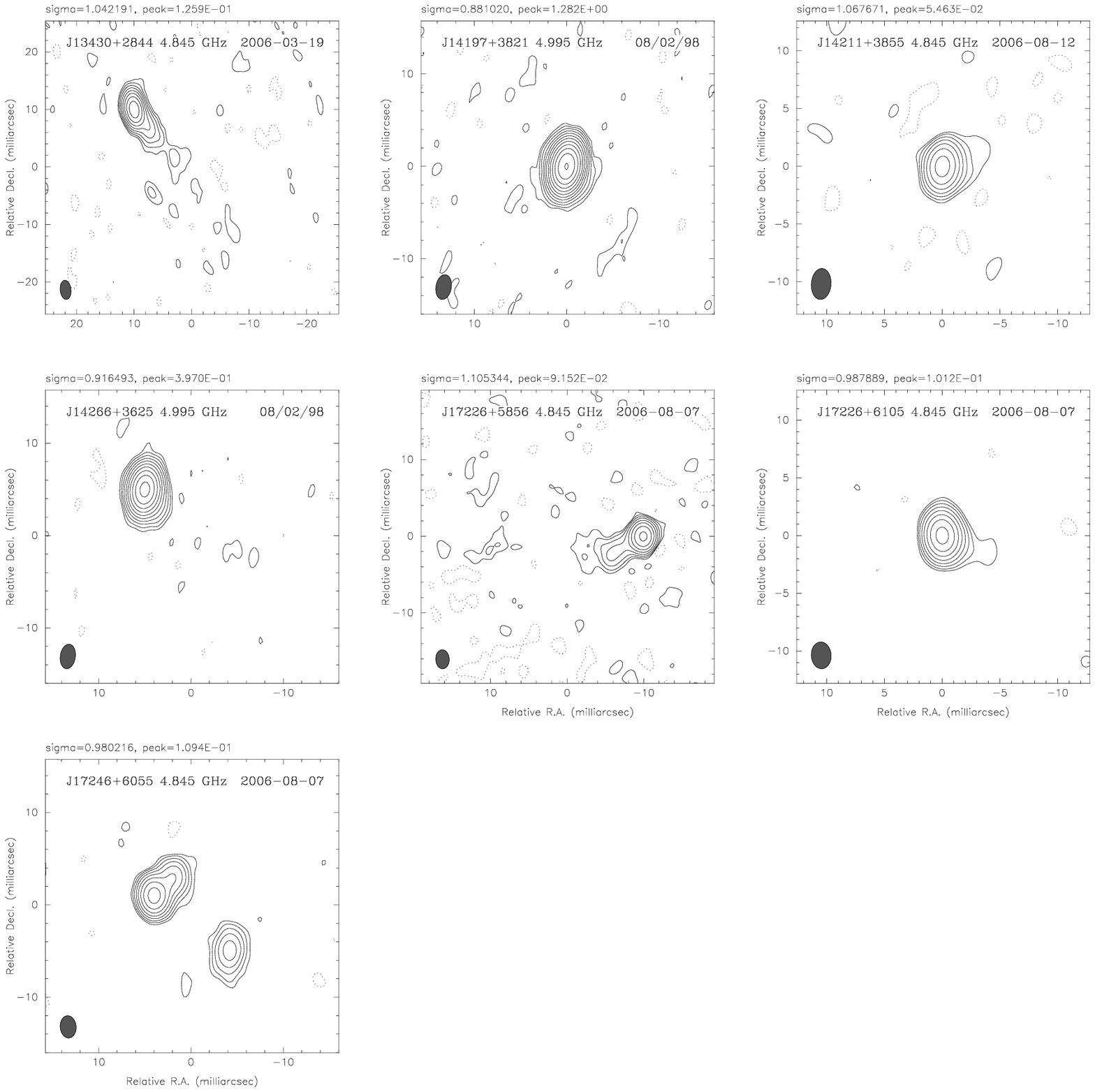}
\caption{Images from VIPS and MOJAVE for 7 sources identified as
plausible EGRET candidates.}
\label{maybe2}
\end{figure}

In some cases multiple VIPS sources are assigned to a single EGRET
source.  Since some EGRET sources may be in fact composites of
multiple blazars we have not made any distinction between sources with
multiple identifications and those with a single identification.  For
6 bright sources otherwise meeting the VIPS source criteria, MOJAVE
observations are available.  In Fig.~\ref{likely1} we show the 5 GHz
VIPS images or 15 GHz MOJAVE images for all 12 sources identified with
EGRET blazars with high confidence (the ``likely'' candidates).  In
Figs.~\ref{maybe1}, and \ref{maybe2} we show the images for 19 sources
identified as plausible counterparts.  For completeness we include 15
GHz MOJAVE images (Kellermann et al. 2004) for those 6 VIPS sources
that were not imaged at 5 GHz.  One should remember that due to the
large range in redshifts, the individual rest-frame emitted
frequencies can range over a factor up to 3.2.

%

\section{Morphological comparison}

While a number of radio-bright EGRET sources
distributed over the sky have been imaged with VLBI (see references 
discussed in \S 1),
only in the VIPS region has the full EGRET candidate sample been
uniformly imaged down to a flux density limit of 85 mJy at 8 GHz. In this
section we examine the VLBI-scale properties of the 31 EGRET candidate
blazars in this uniform sample, and for the 25 sources imaged as
part of VIPS, we compare them to the properties
of the VIPS survey of 1127 sources.

\subsection{Source Classes}

The classification of the EGRET candidates in VIPS is 8 PS (point
sources), 3 SJET (short jets), and 14 LJET (long jets) according to
the automatic classification scheme of Helmboldt et al. (2007a; see
Table 2).  There are no CSO (Compact Symmetric Object) candidates or
CPLX (complex) sources.  Applying the same classification rules to the 6
MOJAVE sources adds 5 LJET and 1 SJET for totals of 8 (25\%) PS, 4
(13\%) SJET, and 19 (61\%) LJET.  This can be compared to the
population in VIPS as a whole which is 25\% PS, 22\% SJET, 43\% LJET,
9\% CSO and 2\% CPLX.  If EGRET candidates were drawn from the VIPS
population at random then we would expect 8 $\pm$ 3 PS, 7 $\pm$ 3
SJET, 13 $\pm$ 4 LJET, and 3 $\pm$ 2 CSO and COMPLEX.  While the
number of PS and SJET sources are roughly as expected we find a
marginally significant lack of CSOs, and a marginally significant excess of
sources with long jets (LJET) on the parsec-scale.

\subsection{Polarization}

Polarization is detected from all 6 of the MOJAVE sources and from 9
of the 25 VIPS sources for a total of 48 $\pm$ 12\%.  This is higher
than the average of 36 $\pm$ 4\% in VIPS overall (Helmboldt et
al. 2007a), and the 41 $\pm$ 4\% average for LJETs, though the numbers
of sources involved are too small to make a definitive statment.
Similarly a detailed examination of the polarization properties is
handicapped by small number statistics and the fact that we have a mix
of 5 GHz measurements for VIPS sources and 15 GHz measurements for
MOJAVE sources.  Lister \& Homan (2005) found that that the jets (but
not the cores) of EGRET blazars tended to have nearly twice the level
of fractional polarization compared to MOJAVE sources not detected
by EGRET, and
are also more luminous.  They attribute the difference to a higher
mean Doppler factor for the EGRET blazars.

\subsection{Core Fraction}

\begin{figure}
\plotone{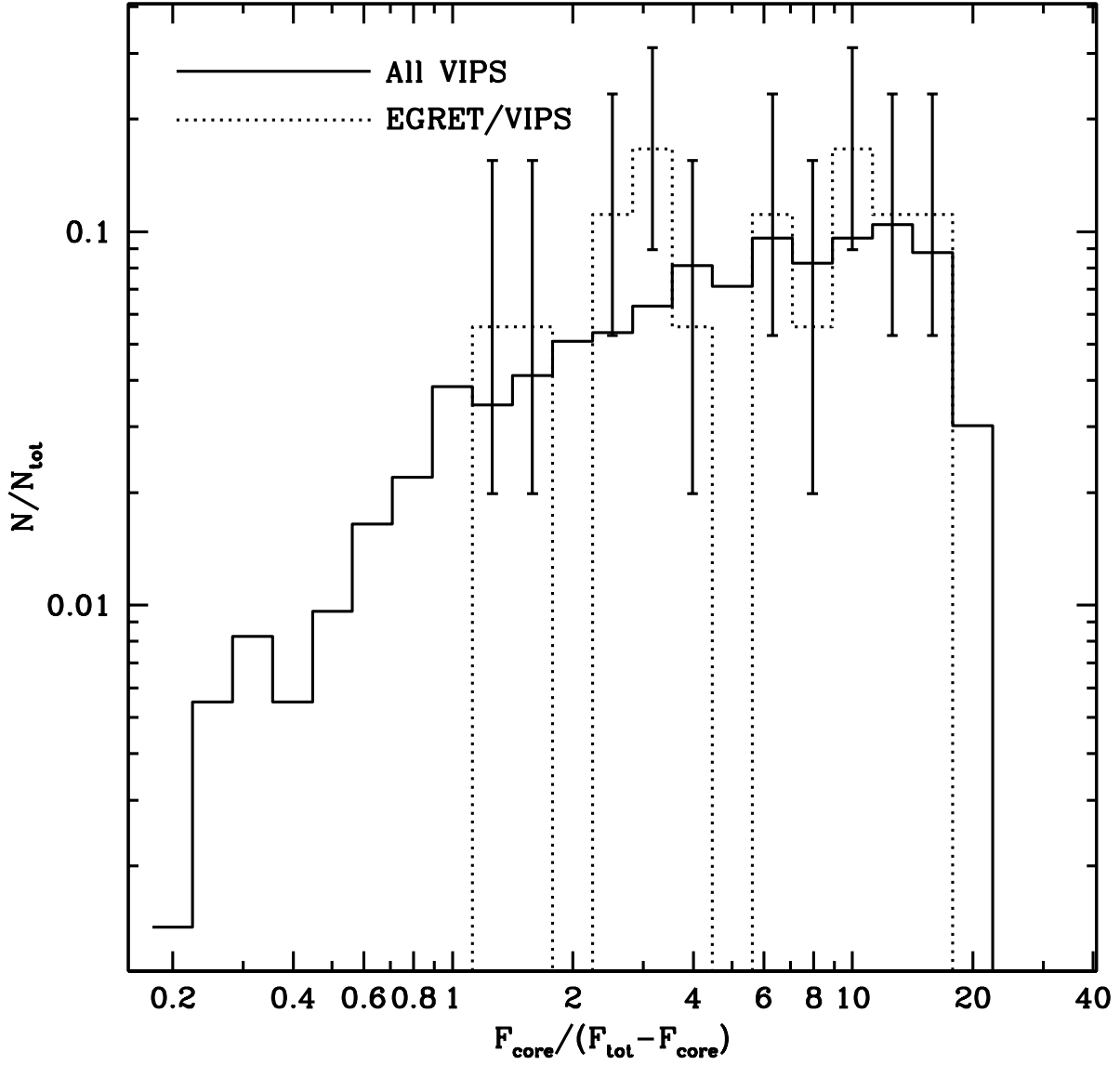}
\caption{Distribution of the 5 GHz ratio of core to jet flux density: $R = F_{core}/(F_{tot} - F_{core})$ where
$F_{tot}$ is the total flux density at 5 GHz, for EGRET candidates
and for VIPS as a whole.  No k-correction has been made (see text).}
\label{cfrac}
\end{figure}

In Fig.~\ref{cfrac} we plot the distribution of core-to-jet ratio
(defined as the ratio of flux density at 5 GHz in the VLBI core compared to the
jet emission from the source at 5 GHz) for the EGRET candidates against the
distribution for the VIPS survey as a whole.  No k-corrections have been
applied, even though the difference in spectral index between the core 
and the extended emission will result in a change in the ratio with
source redshift.  Given the similar range in redshifts, and the core 
dominated nature of the sources, the corrections should be modest and
similar for both EGRET-detected blazars and VIPS sources as a whole.
From the histograms, the
EGRET sources appear to be biased toward more core-dominated
systems.  The K-S test probability that the two distributions were
drawn from the same parent distribution is 4\%.  The core flux density 
for the EGRET blazars can be found in Table 3.

\subsection{Brightness Temperature}

We have obtained the brightness temperatures from automatic modelfits
to source components as described in Helmboldt et al. (2007a).  For
sources with two or fewer components we have further refined the our
modelfitting procedure by fitting to the visibility data directly.
Only image plane modelfitting was carried out in Helmboldt et
al. (2007a) due to the tendency for the automatic visibility
modelfitting to go awry for complicated sources.  The minimum
observable size for each source was calculated using equation (2) from
Kovalev et al. (2005), where we computed the SNR of each core using
the core flux density, the rms measured from the 5 GHz image, and a
beam FWHM of 3 mas, the largest dimension of our restoring beam.  For
those sources where the estimated core size is less than this minimum
size ($\sim$10\% of all VIPS sources), we used the minimum size to compute
$T_B$.  In Table 3 we provide the core sizes, core $T_B$, brightest
jet component $T_B$, $\delta_PA$, and jet opening half angle for the
VIPS blazars.  For sources where the estimated core FWHM is less than
the minimum, the core diameter is preceded by a '$<$' and the core
temperature is preceded by a '$>$'.  
In all cases we compute the observed brightness temperatures, without
any correction to the source rest frame.

\begin{figure}
\plotone{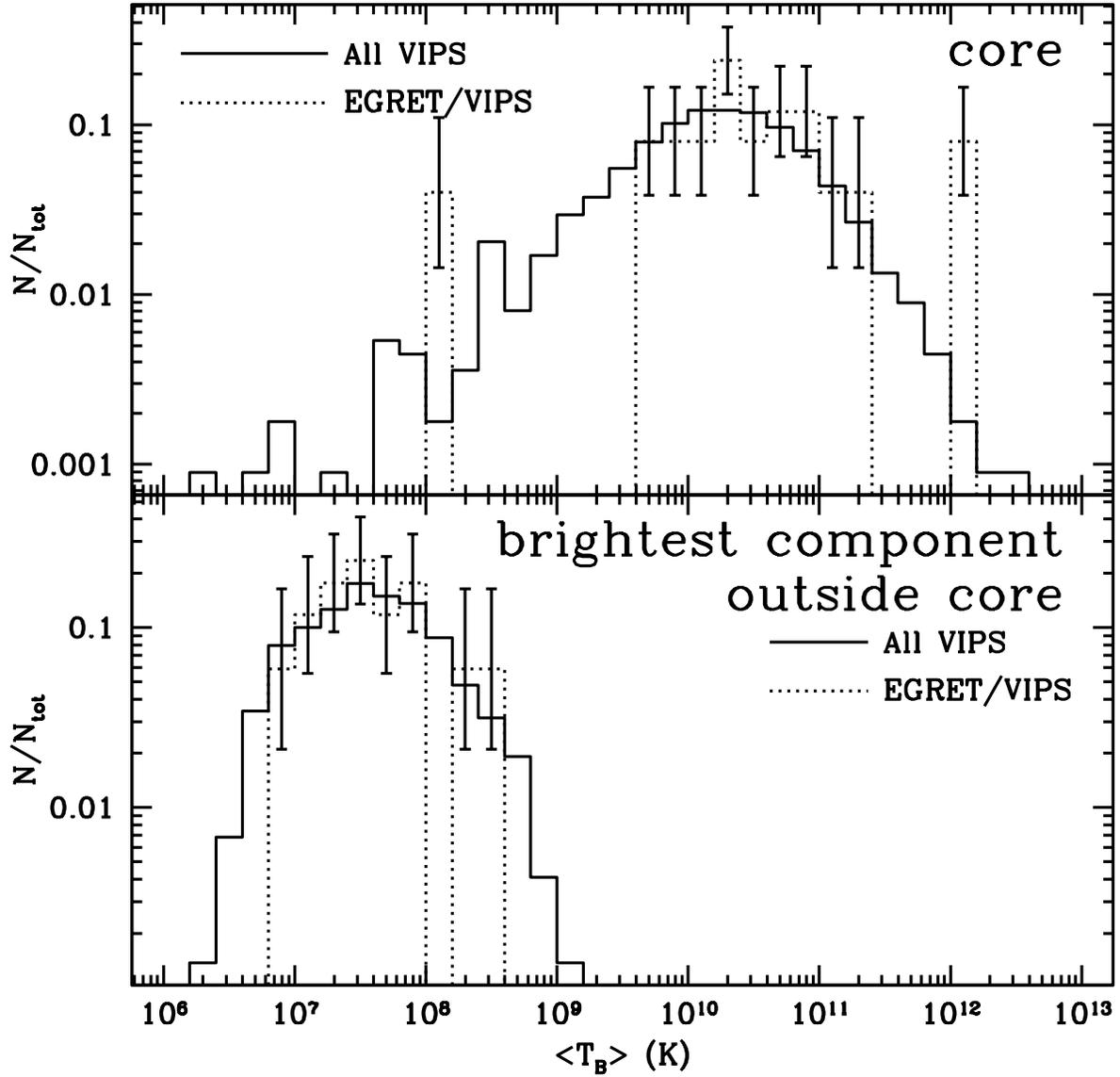}
\caption{Distribution of the brightness temperature, T$_b$, for EGRET
candidates and for VIPS as a whole.}
\label{btemp}
\end{figure}

There is some indication that the brightness temperature for the cores
of the EGRET candidates may be higher than average (see distributions
in Fig.~\ref{btemp}).  The K-S test
probability for these two distributions coming from the same
parent population is 4\%.  For the jet components
(formally, the brightest jet component) the two distributions
look fairly similar, except that the spread in the EGRET distribution
is smaller (see lower panel of Fig.~\ref{btemp}); the K-S test for this
pair of distributions is 11\%.

\subsection{Jet Opening Angle}

\begin{figure}
\plotone{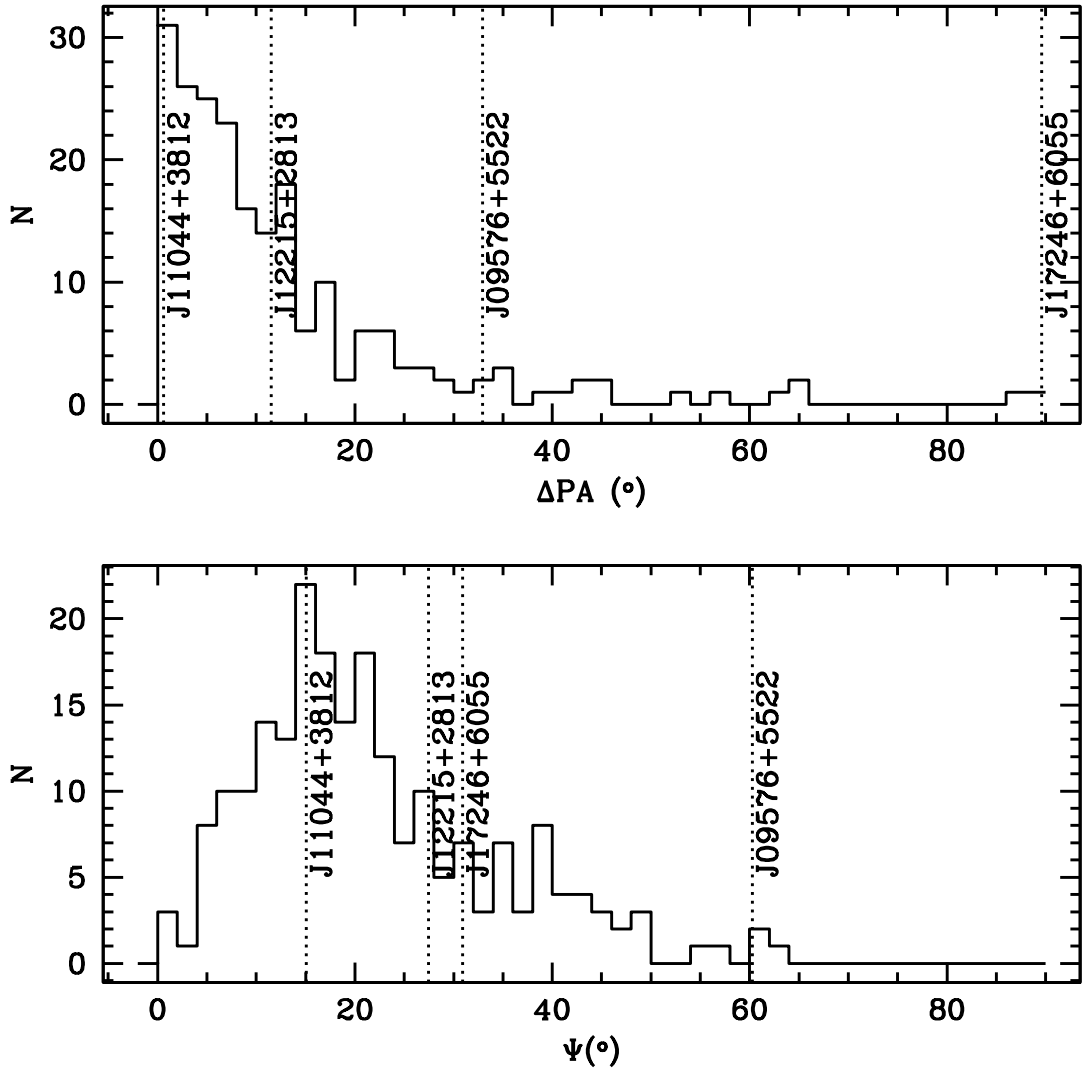}
\caption{Distributions of the opening angle of the jet, and of the 
parsec scale bending of the jet for VIPS as a whole.  The angles for 4 
EGRET sources are marked and labeled.}
\label{angle}
\end{figure}

We have measured a mean opening half-angle by the following 
procedure:  We measure the separation of each jet component from the core along the
jet axis (taken to be a linear fit to the component positions) and the
distance of each component from the jet axis, i.e., x' and y'
positions in a rotated coordinate system with the jet axis along the
x'-axis.  For each component, we measure its extent from its center
along a line perpendicular to the jet axis using the parameters of its
elliptical fit, and then deconvolve this using the extent of the
Gaussian restoring beam along the same line.  The opening half-angle
measured from each component is then taken to be 
$$
\psi = {\rm arctan}[(|y'|+dr)/|x'|]
$$ where $dr$ is the deconvolved Gaussian size perpendicular to the
jet axis.  After measuring this for each jet component, we average
them to get one value.  This is only done for sources with more than 2
total components, (i.e., at least 2 jet components).  This means that
as in the case of jet bending (see \S 3.6), we only have quantitative estimates for
four of the EGRET/VIPS sources (see. Fig.~\ref{angle}).  Though this
is based on very small statistics there is a hint that EGRET/VIPS jets
have unusually large opening angles.  
We find 3 sources have jet opening half-angles exceeding the VIPS median (18.9$^\circ$).
Monte-Carlo simulations of opening angles using the VIPS sample suggests 
that 3 out of 4 of the randomly
sampled values are greater than the median about 30\% of the time.  So,
there's less than a 1 in 3 chance that the EGRET blazars have the same
distribution for the opening angle as the whole VIPS sample.
Looking at the EGRET/MOJAVE and 
EGRET/VIPS sources by eye in Figures 2-4
strengthens this impression.  We see in Figures 2-4 
six sources (J08098+5218, J09576+5522, J12215+2813, 1611+343, J16352+3808, 
J17406+5211) that have opening angles larger than
20 degrees.  This can be compared to the VIPS images for the survey
as a whole which are online\footnote{See http://www.phys.unm.edu/$\sim$gbtaylor/VIPS/}.
In a few of the blazars the emission is rather diffuse
which is why they do not show up in the automatic opening
angle calculation.

\subsection{Jet Bending and Length}

We can also compare the amount of bending in the jet on the parsec
scale for the EGRET candidates compared to VIPS as a whole.  The
amount of bending was calculated by fitting a parabola to the
centroids of Gaussian model components (Helmboldt et al. 2007b).  Again
we are limited to only 4 sources (Fig.~\ref{angle}), and the results
are inconclusive.  In fact the source with the largest reported bend,
J09576+5522, could be much less bent depending on if the morphology is
attributed to a twisting jet, or to a jet with a broad opening angle.
Looking at the EGRET/MOJAVE and EGRET/VIPS sources by eye only
increases the number of highly bent jets by one.  We see in Figures
2-4 4 sources (J12215+2813, J17226+5856, J17246+6055, J17406+5211)
that appear to bend through more than 20 degrees on the parsec scale.

Given the slight preponderance of sources with long jets, we compared
the distribution of jet length in EGRET blazars with VIPS as a whole.
There does not appear to be an appreciable difference between the two
populations.  The K-S probability that the two are drawn from the same
distribution is 34\%.  

\section{Notes on Individual Sources}

\noindent
{\bf J11044+3812 = Mrk 421} A well known BL Lac and also one of the
first sources to be detected in TeV emission (Punch et al. 1992).
Mrk~421 was included in both the CJF (Taylor et al. 1994) 
and 2cm surveys (Kellermann et al. 2004) and has also been 
recently imaged by Giroletti et al. (2006).

\noindent
{\bf J16036+1554} This source is completely unresolved by the 
VLBA observations at 5 GHz (Helmboldt et al. 2007a), and can
be well fit with a single component (the core) with  size of
 $<0.12$ mas.  This results in 
a brightness temperature of $> 1 \times 10^{12}$ K.  Such completely
unresolved sources without any jet components are quite rare - 
there are essentially no
sources this compact in MOJAVE (Kovalev et al. 2005) or CJF
(Taylor et al. 1994).  

\noindent
{\bf J17246+6055} The faint (r band magnitude 21.2) host galaxy does
not yet have a known redshift or spectroscopic classification.  The
parsec-scale radio structure (see Fig.~\ref{maybe2}) is quite unusual
with a core-jet and a fairly compact component to the southeast.  It
is possible that this source could be a compact supermassive binary
black hole, and further VLBI observations are in progress to test
this hypothesis.  Another possible explanation is a jet with a 
sharp bend and wide opening angle, such as Mrk~501 (Giroletti et al. 2004) or 
Mrk~421 (Giroletti et al. 2006; and see Fig.~2).

\section{Discussion}

From an analysis of the properties of the EGRET candidates in \S4, we
find that candidates tend to have high core fractions and high
brightness temperatures.  These facts support models wherein the
gamma-rays are produced in a jet oriented at a small angle to the
line-of-sight (e.g., Dermer et al. 1992).  
The large opening angles seen in 6 out of the 31
sources could also indicate a sight-line close to or within the
beaming cone of the radio emission.  A selection effect of this sort
would happen naturally if the gamma rays are more tightly beamed or
collimated than the radio.  In other words, 
gamma-ray selected sources have higher
Doppler factors than the population of compact sources as a whole. 
This can also explain the observations of higher core fractions, 
higher brightness temperatures, and higher polarization in
gamma-ray blazars compared to the population of AGN as a whole, as 
well as the reports of faster motions.  
An unbiased comparison of jet components
between the EGRET blazars and the VIPS sample at large supports this,
though only 4 of the EGRET/VIPS sources have enough jet components to
permit the unbiased (automatic) analysis so the statistics are poor.

Somewhat more surprising is the evidence in support of long jets,
though this is of marginal significance.  Continuing the line of
discussion above that EGRET blazars have higher Doppler factors due to
small angles to the line-of-sight, the {\it a priori} prediction,
assuming jets of a fixed length and modest velocity, would be for
shorter jets due to projection.  Obtaining longer jets might be
possible if the jets of blazars have intrinsically greater surface
brightness, or are highly Doppler boosted compared to sources in radio
selected samples.  Longer jets in blazars could also be obtained if
there is a bias in the selection.  The redshift distribution for the
VIPS sample is still incomplete, so it is difficult to investigate
possible selection bias at this time.  Given the weakness of the
evidence, especially when comparing the distributions of jet lengths
(\S 3.6), further speculation on this topic is not yet warranted.

\section{Conclusions}

Sources selected on the basis of their $\gamma$-ray flux 
as determined by EGRET 
appear to have higher Doppler factors than the radio-selected
population at large.  In the VIPS survey this selection manifests 
itself in larger 
core fractions, higher core brightness temperatures, increased
polarization, and possibly broader jets.  

GLAST, which will launch in late 2007 or early 2008, will detect
several thousand (Gehrels \& Michelson 1999) of the tens of thousands
of compact, radio-loud, flat-spectrum AGN currently known. 
Confirming the tighter beaming angle for the $\gamma$-rays should
be easily achieved with the improved statistics.  Greater advances
in understanding are anticipated, though 
detailed studies including multi-wavelength
flux monitoring, and multi-frequency, multi-epoch VLBI polarimetry
will be possible only for a small subset of objects.  As we have
shown, the gamma-ray flux density is not simply related to the radio
flux-density, so selecting the brightest radio sources for study may
not maximize the chances of coincidence with strong gamma-ray
emission.  Therefore it may be advantageous to consider selecting
sources with large core fractions, high central brightness
temperatures, and large opening angles, as well as strong radio flux
density.  GLAST could also see new populations of AGN including
Seyfert galaxies, and low power radio galaxies.  For this reason
it is important to provide contemporaneous ground-based monitoring of a range
of source classes, powers, and morphologies.  Comparisons between the parsec-scale radio structures as 
revealed by VLBI observations and the time variable $\gamma$-ray flux,
could discriminate between various models describing the
production of the $\gamma$-ray emission, or reveal that more than
one mechanism is at work.

\noindent
We thank an anonymous referee for constructive suggestions.
The National Radio Astronomy Observatory is a facility of the National
Science Foundation operated under cooperative agreement by Associated
Universities, Inc.  

{}

\begin{deluxetable}{ccrrrrrrrrr}
\tabletypesize{\scriptsize}
\tablecolumns{11}
\tablewidth{0pt}
\tablecaption{EGRET Identifications}
\tablehead{
\colhead{3EG name} & \colhead{VIPS name} & \colhead{S 8.4} & \colhead{$\alpha$} & \colhead{$\Delta TS$} & \colhead{FoM} & \colhead{Code} & \colhead{Mag} & \colhead{z} & \colhead{Type} & \colhead{In CGRaBS?}  \\
\colhead{(1)} & \colhead{(2)} & \colhead{(3)} & \colhead{(4)} & \colhead{(5)} & \colhead{(6)} & \colhead{(7)} & \colhead{(8)} & \colhead{(9)} & \colhead{(10)} & \colhead{(11)}  \\}

\startdata
J0743+5447  &  J0742+5444    &    142.8  &  -0.387  &   0.517  &   0.502  & !        &   17.3  &   0.723 &  FSRQ \\
J0808+5114  &  J0807+5117    &    358.8  &   0.280  &   0.359  &   7.241  & !!!      &   17.6  &   1.136 &  FSRQ      & CGRaBS \\
J0808+5114  &  J0809+5218    &    154.2  &  -0.103  &   2.398  &   1.186  & !!!      &   14.5  &   0.138 &  BLL \\
J0829+2413  &  J0830+2410    &    793.7  &   0.027  &   0.701  &   7.009  & !!!      &   16.3  &   0.940 &  FSRQ      & CGRaBS \\
J0853+1941  &  J0854+2006    &   3414.9  &   0.443  &   1.519  &   7.872  & !!!      &   15.6  &   0.306 &  BLL       & CGRaBS \\
J0917+4427  &  J0920+4441    &   1368.3  &   0.153  &   7.051  &   0.470  & !        &   16.5  &   2.190 &  FSRQ      & CGRaBS \\
J0952+5501  &  J0957+5522    &   1498.9  &  -0.410  &   3.839  &   0.427  & !        &   16.8  &   0.896 &  FSRQ \\
J1052+5718  &  J1058+5628    &    189.8  &  -0.115  &   5.490  &   0.387  & !        &   14.6  &   0.144 &  BLL       & CGRaBS \\
J1104+3809  &  J1104+3812    &    631.6  &  -0.109  &   1.126  &   6.582  & !!!      &   13.0  &   0.031 &  BLL       & CGRaBS \\
J1200+2847  &  J1159+2914    &   1233.3  &  -0.286  &   4.034  &   0.829  & !        &   17.4  &   0.729 &  FSRQ      & CGRaBS \\
J1222+2841  &  J1221+2813    &   1055.5  &   0.194  &   8.313  &   0.280  & !        &   14.2  &   0.102 &  BLL       & CGRaBS \\
J1224+2118  &  J1224+2122    &   1073.9  &  -0.383  &   1.445  &   1.522  & !!!      &   16.0  &   0.435 &  FSRQ \\
J1227+4302  &  J1221+4411    &    435.2  &  -0.149  &   5.239  &   0.456  & !        &   17.8  &   1.344 &  FSRQ      & CGRaBS \\
J1227+4302  &  J1224+4335    &    220.8  &  -0.200  &   1.567  &   1.133  & !!!      & \nodata  &   1.075 &  FSRQ \\
J1227+4302  &  J1226+4340    &    145.1  &  -0.141  &   1.367  &   0.866  & !        &   18.5  &   1.999 &  FSRQ \\
J1323+2200  &  J1321+2216    &    323.6  &   0.010  &   5.353  &   0.509  & !        &   19.4  &   0.943 &  FSRQ      & CGRaBS \\
J1323+2200  &  J1322+2148    &    147.3  &  -0.288  &   3.400  &   0.261  & !        &   19.1  &   1.680 &  FSRQ \\
J1323+2200  &  J1327+2210    &    984.0  &   0.073  &   6.547  &   0.381  & !        &   18.8  &   1.400 &  FSRQ      & CGRaBS \\
J1329+1708  &  J1331+1712    &    120.7  &  -0.233  &   1.454  &   0.425  & !        &   17.2  & \nodata &  BLL \\
J1329+1708  &  J1333+1649    &    483.8  &   0.106  &   3.072  &   1.982  & !!!      &   16.4  &   2.097 &  FSRQ      & CGRaBS \\
J1347+2932  &  J1343+2844    &    192.0  &  -0.127  &   5.069  &   0.336  & !        &   16.5  &   0.908 &  FSRQ \\
J1424+3734  &  J1419+3821    &    775.8  &   0.120  &   5.408  &   0.840  & !        &   19.2  &   1.831 &  FSRQ      & CGRaBS \\
J1424+3734  &  J1421+3855    &    132.1  &   0.228  &   4.478  &   0.441  & !        &   17.4  &   0.489 &  FSRQ \\
J1424+3734  &  J1426+3625    &    613.0  &   0.201  &   6.357  &   0.428  & !        &   20.3  &   1.091 &  FSRQ      & CGRaBS \\
J1605+1553  &  J1603+1554    &    256.9  &   0.524  &   2.483  &   4.280  & !!!      &   12.1  &   0.109 &  FSRQ      & CGRaBS \\
J1614+3424  &  J1613+3412    &   3088.1  &  -0.165  &   2.313  &   1.984  & !!!      &   17.1  &   1.397 &  FSRQ      & CGRaBS \\
J1635+3813  &  J1635+3808    &   2403.9  &  -0.085  &   1.488  &   3.599  & !!!      &   17.2  &   1.814 &  FSRQ      & CGRaBS \\
J1733+6017  &  J1722+5856    &    328.0  &   0.189  &   6.953  &   0.264  & !        &   19.0  &   1.979 &  FSRQ      & CGRaBS \\
J1733+6017  &  J1722+6105    &    203.2  &   0.141  &   3.932  &   0.911  & !        &   19.4  &   2.058 &  FSRQ      & CGRaBS \\
J1733+6017  &  J1724+6055    &    166.0  &  -0.042  &   2.340  &   0.827  & !        & \nodata  &\nodata \\
J1738+5203  &  J1740+5211    &   1347.2  &   0.275  &   0.314  &  11.149  & !!!      &   16.7  &   1.379 &  FSRQ      & CGRaBS \\ 
\enddata

\tablecomments{Col.\ (1): 3EG name based on EGRET coordinates.  
Col.\ (2): VIPS source name based on CLASS coordinates.
Col.\ (3): 8.4 GHz flux density from CLASS.
Col.\ (4): $\alpha$ is the spectral index ($S \sim \nu^\alpha$) between 
NVSS (Condon et al. at 1.4 GHz and CLASS at 8.4 GHz,
Col.\ (5): ``$\Delta$TS'' is a measure of the positional correspondence of the radio
source to the peak of the gamma-ray flux.  Lower values indicate a better
positional correspondence (see Sowards-Emmerd et al. 2003).
Col.\ (6): ``FoM'' is the figure of merit (Sowards-Emmerd et al. 2003). 
Col.\ (7): ``Code'' is a classification of the same type as in Sowards-Emmerd et al.
2003: sources with FoM $>$ 1.0 were deemed ``likely'' counterparts (denoted
here by ``!!!'') and sources with 0.25 $<$ FoM $<$ 1.0 were deemed 
plausible counterparts (denoted here by ``!'').
Col.\ (8):  The optical magnitude of the host galaxy.
Col.\ (9):  The redshift from V\'{e}ron-Cetty \& V\'{e}ron (2006).
Col.\ (10): Source classification, where FSRQ = Flat Spectrum 
Radio Quasar, and BLL = BL Lac object.
Col.\ (11):  An indication if the source is included in the 
CBRaBS sample (Healey et al. 2007b, in prep).}
\label{egret}
\end{deluxetable}

\textwidth = 8.0truein
\begin{deluxetable}{r@{$\;\;$}c@{$\;\;$}c@{$\;\;$}c@{$\;\;$}c@{$\;\;$}r@{$\;\;$}r@{$\;\;$}r@{$\;\;$}c@{$\;\;$}c@{$\;\;$}c@{$\;\;$}r@{$\;\;$}r@{$\;\;$}r@{$\;\;$}r@{$\;\;$}r@{$\;\;$}r@{$\;\;$}r@{$\;\;$}r}
\rotate
\tabletypesize{\scriptsize}
\tablecolumns{15}
\tablewidth{0pt}
\tablecaption{VIPS Source Properties}
\tablehead{
\colhead{} & \colhead{} & \colhead{$\alpha$ (J2000)} & \colhead{$\delta$ (J2000)} & \colhead{} & \colhead{$F_{8.5}$} & \colhead{$F_{5}$} & \colhead{$F_{5,max}$} & \colhead{rms$_{5}$} & \colhead{} & \colhead{} & \colhead{} & \colhead{$\overline{R}$} & \colhead{$D_{max}$} & \colhead{$PA_{jet}$} \\
\colhead{} & \colhead{Name} & \colhead{($^{h}\:^{m}\:^{s}$)} & \colhead{($^{\circ}\:^{'}\:^{''}$)} & \colhead{UT Date} & \colhead{(mJy)} & \colhead{(mJy)} & \colhead{(mJy/beam)} & \colhead{(mJy/beam)} & \colhead{$N_{GC}$} & \colhead{T$_{\mbox{\scriptsize a}}$} & \colhead{T$_{\mbox{\scriptsize e}}$} & \colhead{(mas)} & \colhead{(mas)} & \colhead{($^{\circ}$)} \\
\colhead{(1)} & \colhead{(2)} & \colhead{(3)} & \colhead{(4)} & \colhead{(5)} & \colhead{(6)} & \colhead{(7)} & \colhead{(8)} & \colhead{(9)} & \colhead{(10)} & \colhead{(11)} & \colhead{(12)} & \colhead{(13)} & \colhead{(14)} & \colhead{(15)}\\
\vspace{-0.3cm}}
\startdata
   72 & J08070+5117 & 08:07:01.0133 & +51:17:38.670 & 2006-05-31 &    358.8 &    175.9 &     84.0 &     0.23 &        1 &       PS &  \nodata &      1.3 &  \nodata &  \nodata \\ 
   78 & J08098+5218 & 08:09:49.1899 & +52:18:58.252 & 2006-05-31 &    154.2 &    125.6 &     56.1 &     0.25 &        2 &     LJET &  \nodata &      3.4 &      6.7 &     20.2 \\ 
  402 & J11044+3812 & 11:04:27.3145 & +38:12:31.794 & 1999-11-21 &    631.6 &    421.4 &    372.0 &     0.34 &        3 &     LJET &  \nodata &      9.6 &     23.8 &  $-$42.3 \\ 
  554 & J12248+4335 & 12:24:51.5074 & +43:35:19.276 & 2006-05-27 &    220.8 &    173.5 &     98.2 &     0.21 &        2 &     LJET &  \nodata &      4.0 &      8.0 &     13.0 \\ 
  683 & J13335+1649 & 13:33:35.7819 & +16:49:04.027 & 2006-03-19 &    483.8 &    442.2 &    310.0 &     0.28 &        2 &     LJET &  \nodata &     12.8 &     25.5 &     18.6 \\ 
  937 & J16036+1554 & 16:03:38.0650 & +15:54:02.378 & 2006-04-03 &    256.9 &    348.3 &    347.5 &     0.20 &        1 &       PS &  \nodata &      1.2 &  \nodata &  \nodata \\ 
  994 & J16352+3808 & 16:35:15.4931 & +38:08:04.497 & 1999-11-21 &   2423.3 &   2201.9 &   1240.0 &     0.46 &        2 &     SJET &  \nodata &      2.4 &      4.8 &     86.2 \\ 
 1102 & J17406+5211 & 17:40:36.9805 & +52:11:43.413 & 1999-11-21 &   1358.5 &    609.4 &    507.4 &     0.30 &        2 &     LJET &  \nodata &      3.2 &      6.5 &      0.3 \\ 
   27 & J07426+5444 & 07:42:39.7904 & +54:44:24.679 & 2006-05-31 &    142.8 &    207.0 &    163.4 &     0.23 &        1 &       PS &  \nodata &      1.2 &  \nodata &  \nodata \\ 
  214 & J09209+4441 & 09:20:58.4599 & +44:41:53.988 & 1996-08-17 &   1368.3 &   1173.5 &    926.0 &     0.71 &        2 &     LJET &  \nodata &      4.2 &      8.5 &  $-$18.6 \\ 
  282 & J09576+5522 & 09:57:38.1837 & +55:22:57.740 & 1999-11-21 &   1498.9 &    563.2 &     80.2 &     1.99 &        4 &     LJET &  \nodata &     19.1 &     55.9 &     19.6 \\ 
  387 & J10586+5628 & 10:58:37.7261 & +56:28:11.180 & 2006-06-19 &    189.8 &    149.4 &     86.1 &     0.22 &        2 &     LJET &  \nodata &      3.5 &      7.0 &  $-$86.6 \\ 
  548 & J12214+4411 & 12:21:27.0450 & +44:11:29.667 & 1996-08-22 &    435.2 &    351.1 &    288.6 &     0.32 &        2 &     LJET &  \nodata &      5.4 &     10.7 &     52.5 \\ 
  550 & J12215+2813 & 12:21:31.6936 & +28:13:58.497 & 2006-02-25 &    893.1 &    401.2 &    149.4 &     0.24 &        6 &     LJET &  \nodata &      9.5 &     25.5 &  $-$44.3 \\ 
  556 & J12269+4340 & 12:26:57.9051 & +43:40:58.438 & 2006-05-27 &    145.1 &     77.8 &     44.0 &     0.23 &        2 &     LJET &  \nodata &      3.2 &      6.4 &   $-$8.2 \\ 
  655 & J13211+2216 & 13:21:11.2041 & +22:16:12.098 & 2006-02-25 &    323.6 &    253.1 &    160.9 &     0.19 &        2 &     SJET &  \nodata &      2.3 &      4.7 &     58.2 \\ 
  656 & J13221+2148 & 13:22:11.3989 & +21:48:12.268 & 2006-02-25 &    147.3 &    171.0 &    131.9 &     0.21 &        1 &       PS &  \nodata &      1.2 &  \nodata &  \nodata \\ 
  679 & J13315+1712 & 13:31:33.4457 & +17:12:50.615 & 2006-02-25 &    120.7 &     66.3 &     41.6 &     0.21 &        2 &     SJET &  \nodata &      2.5 &      5.0 & $-$113.4 \\ 
  697 & J13430+2844 & 13:43:00.1805 & +28:44:07.492 & 2006-03-19 &    192.0 &    153.2 &     99.3 &     0.24 &        2 &     LJET &  \nodata &      3.2 &      6.5 & $-$140.9 \\ 
  754 & J14197+3821 & 14:19:46.6162 & +38:21:48.485 & 1998-02-08 &    775.8 &   1318.1 &   1282.0 &     0.23 &        1 &       PS &  \nodata &      1.1 &  \nodata &  \nodata \\ 
  758 & J14211+3855 & 14:21:06.0337 & +38:55:22.829 & 2006-08-12 &    132.1 &     54.4 &     49.8 &     0.20 &        1 &       PS &  \nodata &      1.1 &  \nodata &  \nodata \\ 
  769 & J14266+3625 & 14:26:37.0861 & +36:25:09.576 & 1998-02-08 &    613.0 &    432.9 &    397.0 &     0.22 &        1 &       PS &  \nodata &      1.1 &  \nodata &  \nodata \\ 
 1073 & J17226+5856 & 17:22:36.7292 & +58:56:22.265 & 2006-08-07 &    328.0 &    111.9 &     66.3 &     0.27 &        2 &     LJET &  \nodata &      3.5 &      6.9 &  $-$60.8 \\ 
 1074 & J17226+6105 & 17:22:40.0595 & +61:05:59.802 & 2006-08-07 &    203.2 &    105.8 &     96.6 &     0.19 &        1 &       PS &  \nodata &      1.2 &  \nodata &  \nodata \\ 
 1082 & J17246+6055 & 17:24:41.4172 & +60:55:55.746 & 2006-08-07 &    166.0 &    157.4 &     91.7 &     0.28 &        3 &     LJET &  \nodata &      5.8 &     12.7 & $-$131.2 \\ 
\enddata

\vspace{-0.4cm}
\tablecomments{Col.\ (1): VIPS source number.  Col.\ (2): VIPS source name.  Col.\ (3): Right ascension (J2000).  Col.\ (4): Declination (J2000).  Col.\ (5): UT date of the observations. Col.\ (6): The flux density at 8.5 GHz from the CLASS survey.  Col.\ (7): The total cleaned flux density from the 5 GHz VLBA map. Col.\ (8): The peak flux density from the 5 GHz VLBA map.  Col.\ (9): The rms noise of the 5 GHz VLBA image.  Col.\ (10): The number of dominant Gaussian components (i.e., that contain more than 95\% of the total flux) fit to the 5 GHz VLBA map (see \S 2.3).  Col.\ (11): The source type derived from the automated Gaussian component classification (see \S 2.3).  Col.\ (12): The source type determined by visual inspection of the I image for sources where the ``by-eye'' and automatic classifications disagree.  Col.\ (13): The mean radius (i.e., mean distances from the mean component position) for the ensemble of dominant Gaussian components.  Col.\ (14): The maximum separation among the dominant Gaussian components.  Col.\ (15): The jet position angle (measured from north through east).} 
\label{props}
\end{deluxetable}

\textwidth = 8.0truein
\begin{deluxetable}{lccccccc}
\tabletypesize{\scriptsize}
\tablecolumns{8}
\tablewidth{0pc}
\tablecaption{Brightness Temperatures of VIPS Blazars}
\tablehead{\colhead{} & \colhead{} & \colhead{$F_{5,core}$} & \colhead{$D_{core}$} & \colhead{$T_{B,core}$} & \colhead{$T_{B,jet}$} & \colhead{$\Delta$PA} & \colhead{$\Psi$} \\
\colhead{} & \colhead{Name} & \colhead{(mJy)} & \colhead{(mas)} & \colhead{(K)} & \colhead{(K)} & \colhead{$(^{\circ})$} & \colhead{$(^{\circ})$} \\
\colhead{(1)} & \colhead{(2)} & \colhead{(3)} & \colhead{(4)} & \colhead{(5)} & \colhead{(6)} & \colhead{(7)} & \colhead{(8)}}
\startdata
  72 & J08070+5117 & 175.9 &  0.942 &  9.70$\times 10^{9}$ & \nodata & \nodata & \nodata \\ 
  78 & J08098+5218 & 108.9 &  1.050 &  4.83$\times 10^{9}$ & 1.56$\times 10^{7}$ & \nodata & \nodata \\ 
 402 & J11044+3812 & 370.2 &  0.291 &  2.13$\times 10^{11}$ & 3.59$\times 10^{7}$ & -0.62 & 15.07 \\ 
 554 & J12248+4335 & 156.9 &  0.653 &  1.80$\times 10^{10}$ & 3.37$\times 10^{7}$ & \nodata & \nodata \\ 
 683 & J13335+1649 & 417.9 &  0.538 &  7.05$\times 10^{10}$ & 9.97$\times 10^{6}$ & \nodata & \nodata \\ 
 937 & J16036+1554 & 348.3 & $<$0.121 & $>$1.16$\times 10^{12}$ & \nodata & \nodata & \nodata \\ 
 994 & J16352+3808 & 2013.7 &  1.999 &  2.47$\times 10^{10}$ & 2.74$\times 10^{8}$ & \nodata & \nodata \\ 
1102 & J17406+5211 & 574.7 &  0.700 &  5.75$\times 10^{10}$ & 4.09$\times 10^{7}$ & \nodata & \nodata \\ 
  27 & J07426+5444 & 207.0 &  0.439 &  5.26$\times 10^{10}$ & \nodata & \nodata & \nodata \\ 
 214 & J09209+4441 & 1080.4 &  0.785 &  8.58$\times 10^{10}$ & 5.19$\times 10^{7}$ & \nodata & \nodata \\ 
 282 & J09576+5522 & 127.3 &  7.206 &  1.20$\times 10^{8}$ & 7.96$\times 10^{7}$ & -32.94 & 60.29 \\ 
 387 & J10586+5628 & 129.6 &  0.817 &  9.49$\times 10^{9}$ & 2.10$\times 10^{7}$ & \nodata & \nodata \\ 
 548 & J12214+4411 & 303.6 &  0.528 &  5.33$\times 10^{10}$ & 2.63$\times 10^{7}$ & \nodata & \nodata \\ 
 550 & J12215+2813 & 177.5 &  0.614 &  2.30$\times 10^{10}$ & 1.62$\times 10^{8}$ & -11.50 & 27.45 \\ 
 556 & J12269+4340 & 72.9 &  0.578 &  1.07$\times 10^{10}$ & 1.54$\times 10^{7}$ & \nodata & \nodata \\ 
 655 & J13211+2216 & 227.7 &  0.780 &  1.83$\times 10^{10}$ & 6.51$\times 10^{7}$ & \nodata & \nodata \\ 
 656 & J13221+2148 & 171.0 &  0.465 &  3.87$\times 10^{10}$ & \nodata & \nodata & \nodata \\ 
 679 & J13315+1712 & 61.0 &  0.788 &  4.80$\times 10^{9}$ & 1.86$\times 10^{7}$ & \nodata & \nodata \\ 
 697 & J13430+2844 & 137.3 &  0.602 &  1.85$\times 10^{10}$ & 3.20$\times 10^{7}$ & \nodata & \nodata \\ 
 754 & J14197+3821 & 1318.1 &  0.242 &  1.10$\times 10^{12}$ & \nodata & \nodata & \nodata \\ 
 758 & J14211+3855 & 54.4 & $<$0.304 & $>$2.88$\times 10^{10}$ & \nodata & \nodata & \nodata \\ 
 769 & J14266+3625 & 432.9 &  0.503 &  8.36$\times 10^{10}$ & \nodata & \nodata & \nodata \\ 
1073 & J17226+5856 & 97.2 &  0.574 &  1.44$\times 10^{10}$ & 2.26$\times 10^{7}$ & \nodata & \nodata \\ 
1074 & J17226+6105 & 105.8 &  0.224 &  1.03$\times 10^{11}$ & \nodata & \nodata & \nodata \\ 
1082 & J17246+6055 & 121.5 &  0.604 &  1.63$\times 10^{10}$ & 6.71$\times 10^{7}$ & -89.59 & 30.91 \\ 
\enddata

\tablecomments{Col. (1): VIPS source number.  Col. (2): VIPS source name.  Col. (3): The 5 GHz flux density of the core.  Col. (4): The FWHM of the core component estimated using a circular Gaussian fit to the visibility data.  Col. (5): The mean brightness temperature of the core.  Col. (6): The mean brightness temperature of the brightest component outside the core.  Col. (7): The difference between the outer and inner position angles of the jet for sources with more than 2 jet components.  Col. (8): The jet opening half-angle for sources with more than 2 jet components.}
\end{deluxetable}

\textwidth = 7.2truein

\end{document}